# Radiation enhancement and radiation suppression by a left-handed metamaterial

A. D. Boardman and K. Marinov[1]

Photonics and Nonlinear Science Group, Joule Physics Laboratory, University of Salford,

Salford M5 4WT, UK

**Abstract**

The perfect lens property of a dispersive and lossy left-handed metamaterial (LHM) disk is exploited to superimpose a source of electromagnetic radiation onto its mirror image, formed as a result of reflection from a perfect electric conductor (PEC) or a perfect magnetic conductor (PMC). The superposition of a vertical wire-dipole antenna with its PEC-image results in an increase of the radiation resistance of the antenna compared to that of an antenna emitting in free space. On the other hand, if the same antenna is coupled to a PMC-image it is shown that the result is the formation of a non-radiating configuration. The finite-difference time-domain (FDTD) analysis is performed and this allows a detailed characterization of the systems. It is shown that the non-radiating system allows relatively large amounts of electromagnetic energy to be stored in the LHM-disk and that is indicative of strong electromagnetic fields inside the material. This property is employed in a second-harmonic generation (SHG) process and the potential of a non-radiating configuration as an efficient nonlinear device is demonstrated.

---

[1] E-mail: k.marinov@salford.ac.uk

# 1. Introduction

Non-radiating configurations [1-7] are time-dependent charge-current distributions that do not radiate electromagnetic energy. Trivial systems – e.g. an antenna in a Faraday cage – are of no interest. The early work on the subject ([1] and the references therein) appeared in the early days of atomic physics and it was related to the question of stability of atoms. The general conditions under which a charge-current distribution does not radiate electromagnetic energy have been derived already [1, 2]. It has been shown that the fundamental non-uniqueness of the solution of the inverse source problem – the problem of reconstructing the mathematical form of a wave-source from the fields it generates *outside* its own volume – is, in fact, due to the existence of non-radiating components [3]. Therefore, additional information, in the form of a constraint, is necessary in order to reconstruct a source from the field it generates. Using the "minimum energy" constraint, source decomposition into a sum of non-radiating and purely radiating components has been demonstrated. In addition the question of possible optimization of antenna performance, by manipulating its non-radiating component, has been also discussed [4].

Non-radiating systems can be constructed by combining toroidal and supertoroidal solenoids with electric and magnetic dipoles. The ability of some of these systems to generate time-dependent electromagnetic potentials in the absence of electromagnetic fields has been investigated [5-7]. Such systems can be used as detectors for the permittivity of the ambient matter [8, 9].

The "superlens" property of the LHMs [10] can be exploited to create a non-radiating system [11-14]. A frequency-domain analysis [11-13], carried out analytically, can give a detailed picture of the distribution of both the propagating and the evanescent electromagnetic field components. On the other hand, FDTD embodies the advantage that features too difficult (or impossible) to deal with analytically can be included in the model. Besides, it also yields a system analysis in terms of observable quantities like input impedance, radiation resistance and loss resistance [14].

An important outstanding question concerns the possible applications that non-radiating systems might have. In this connection, it should be emphasized that it is not the mere absence of radiation that attracts the attention. Rather it is the unique field configuration, formed "inside" any non-radiating system. It is clear, therefore, that suitable system designs are needed in order to assess, on a practical level, the potential of these systems for applications and device construction. One way of addressing this issue is to realize that the condition for absence of radiation [1, 2] is a relation between the system parameters and the parameters of the ambient environment. Hence, any deviation from this condition stimulates radiation. This possibility has been explored, partially, [8, 9, 14], where the application of non-radiating systems as sensitive detectors has been demonstrated.

Here radiating and non-radiating systems, based on LHMs are considered. It is shown that the radiation resistance of a wire-dipole antenna can be increased or decreased, by using the "perfect lens" property of the LHMs to effectively superimpose the antenna onto its mirror image. The latter is formed as a result of a reflection from a PEC, or PMC boundary. Also, by using a non-radiating system, the efficiency of a second harmonic generation process, resulting from the presence of diodes introduced into the LHM [15], is investigated and the outcome shows the advantage provided by the non-radiating system.

## 2. System design and principle of operation

Figure 1 illustrates how LHMs can be used to control the radiation properties of a vertical wire-dipole antenna. Basic image theory (see e.g. [16]) states that the electromagnetic field created by a source emitting near a PEC, or PMC, mirror can be presented as a superposition of the field of the source and the field created by its image [Figs. 1 (a) and 1 (b)]. Without loss of generality vertical wire-dipoles with azimuthal symmetry are considered. Horizontal dipoles can be used as well [12]. The mirror images of vertical dipoles are dipoles of the same magnitude and in the case of a PEC point in the same direction, whereas for a PMC their direction is reversed. If the antenna is relatively far (on a wavelength scale) from the boundary, the presence of the latter affects the radiation pattern of the antenna but has very little effect on the radiation resistance. The total emitted power is that of the same antenna emitting in free space. On the other hand, if the antenna is close to the boundary the total emitted power must take into account the presence of the boundary. Hence, if an infinitesimal electric dipole is placed on a PMC-surface this action will, effectively, superimpose the dipole and its image (Fig. 1(c)). In this case no power is emitted at all. In contrast, placing the same dipole on a PEC-surface leads to doubling the antenna radiation resistance as compared with that of the same antenna emitting in free space (Fig. 1(d)). Both the situations shown in Fig. 1 (c) and (d) are known from antenna theory. Even though a vertical electric dipole emitting close to a PMC surface (or, equivalently, a horizontal electric dipole emitting close to a PEC-surface) form non-radiating configurations, these particular systems are nevertheless trivial and of no interest, since the fields they create are zero everywhere. On the other hand LHM-driven non-radiating systems are capable of creating large (on a wavelength scale) volumes of non-propagating fields and are likely to find applications. This type of system can be used as a sensitive detector [14].

The LHM permits the creation of a non-radiating property of a vertical dipole on a PMC surface without bringing the dipole into contact with the surface as Fig. 1(e) shows. An LHM disk of thickness $D$ with $\text{Re}(\varepsilon) = -1$ and $\text{Re}(\mu) = -1$ is used here. This situation, shown in Fig 1(e), is equivalent to the situation depicted in Fig. 1(c). The explanation of this outcome derives from the fact that an LHM disk with an effective permittivity and permeability both equal to -1 is, in fact, an electromagnetic *annihilator*. The meaning of the latter term becomes apparent when it is realized that any changes attained by the electromagnetic field as a result of propagation in a slab of free space of thickness *D* will be *undone* (annihilated) by a subsequent propagation in a slab of LHM of the same thickness with an effective refractive index equal to -1. Thus, as far as electromagnetic field is concerned, the antenna is effectively located on the surface. This results in no radiation being emitted, according to Fig. 1(c). At the same time it is clear that a non-propagating electromagnetic field exists in the volume between the antenna and the disk shown in Fig. 1(e).

The same "annihilation" property of the LHM disk produces the equivalence between the systems shown in Fig. 1(d) and Fig. 1(f). In this case coupling the antenna with its image yields an increase of the antenna radiation resistance.

The PMC boundary is a fictitious but useful concept. Physically the effect can be created by placing a pair of identical dipoles fed with $\pi$-out-of-phase voltages in the focal points of the LHM disk [13, 14]. Another way to suppress the radiation is to abandon the PMC in favor of a single horizontal dipole located above a LHM disk sitting on a PEC surface [12]. Note, however, that an actual PMC-like metamaterial has recently been reported [17].

## 3. Results and discussion

The systems shown in Fig. 1(e) and Fig. 1(f) are modeled with the FDTD method [18]. The LHM disk of radius $R_L$ and thickness *D* is the isotropic homogenized outcome of an array of wires and an array of split-ring resonators [19-21]. Such homogenization permits the use of an effective relative permittivity and permeability functions of the material, given by

$$\varepsilon(\omega) = 1 - \frac{\omega_p^2}{\omega(\omega + i\nu)} \tag{1}$$

and

$$\mu(\omega) = 1 + \frac{F\omega^2}{\omega_0^2 - \omega^2 - i\omega\gamma}, \tag{2}$$

respectively. An isotropic metamaterial can be built by arranging the split-ring resonators and the wires in a cubic lattice [20]. In (1) and (2) $\omega$ is the excitation angular frequency, $\omega_p$ is the effective plasma frequency, $\omega_0$ is the resonant frequency, $\nu$ and $\gamma$ are the loss parameters and $F$ is the filling factor. A thin-wire model [22] has been employed for the center-fed wire-dipole antenna of radius $r_0 = 0.9$ mm and length $L_D$. This model has the advantage that it makes no assumptions for the current distribution along the antenna. With $\omega_0/2\pi = 836.7\ MHz$, $\omega_p = 1.414\ GHz$ and $F = 0.6$, the real parts of both $\varepsilon$ and $\mu$ are both equal to -1 at the operating frequency $\omega/2\pi = 1\ GHz$. The corresponding free-space wavelength is $\lambda = 0.299\ m$. The loss parameters are set to $\gamma/2\pi = 5\ MHz$ and $\nu = 0$. The radius of the disk is $R_L = 3.75\lambda = 1.12\ m$. The electromagnetic field components that are not identically zero are $E_r$, $H_\varphi$ and $E_z$. Cylindrical system of coordinates is used. Details of the FDTD model of the dispersive and lossy LHM, with $\varepsilon$ and $\mu$ given by (1) and (2) can be found in [23]. The accuracy of the simulation is controlled by monitoring the extent to which the energy conservation law

$$P_{in} = P_{rad} + P_{loss}, \qquad (3)$$

is satisfied. In (3), $P_{in}$ is the input power supplied to the antenna, $P_{rad}$ is the radiated power and $P_{loss}$ is the power loss.

Figure 2 shows the input resistance $R_{in}$, the radiation resistance $R_{rad}$ and the loss resistance $R_{loss}$ for the non-radiating system shown in Fig. 1(e). Provided that the antenna length $L_D$ is smaller than the thickness of the LHM slab, the radiation resistance of the system remains a small fraction of the input resistance. This means that the input power is almost entirely absorbed by the material, with only a small amount of it being radiated. As Fig. 2(b) shows, in the absence of the LHM disk, the PMC boundary by itself has practically no impact on the input resistance of the antenna. Note that the distance between the center of the antenna and the PMC boundary is $2D = 1.5\lambda$ and this is sufficiently far away. If the antenna length is greater than the thickness of the LHM disk, then the coupling between the antenna and its image is only partial and the result is that the current distribution in the antenna becomes highly asymmetric.

According to Fig. 2 the system shown in Fig. 1(e) has a nearly zero radiation resistance. If the PMC plate is replaced by a PEC plate (Fig. 1(f)) an increase of the radiation resistance is expected, since in this case the image of the dipole is a dipole pointing in the same direction. Indeed, placing an infinitesimal vertical dipole directly on a PEC surface (Fig. 1(d)) increases the radiation resistance of the dipole by a factor of two. The same upper limit for the relative increase of the radiation resistance of a finite dipole emitting over a LHM disk in contact with a PEC plate is expected. The results for the normalized input and radiation resistance of such a system presented in Fig. 3 show severe limitation by losses in the disk. In fact, for the parameter values chosen, the presence of the LHM disk offers no advantage in comparison with a dipole antenna placed directly on top of the PEC plate (the curve $R_{in}$, contact). Decreasing the thickness of the LHM disk from $D = 0.75\lambda$ to $D = 0.5\lambda$, indeed, produces a higher radiation resistance. However, the increase of the loss resistance is even stronger. Note the decrease of both the normalized input resistance and the normalized radiation resistance for antenna lengths greater than the LHM thickness. This is because the complete superposition of the antenna and its image is no longer possible. The increase of the radiation resistance of a system of two identical emitters, coupled by a slab of an LHM has been reported recently [14]. As shown here this can be explained by the effective superposition of the two sources (or the source and its image) ensured by the LHM slab.

The radiation enhancement and radiation suppression properties demonstrated so far offer a straightforward way to test experimentally the ability of a given sample of an LHM to create a "perfect" image of a realistic radiation source, such as a wire-dipole antenna. Placing the sample on a PEC plate and measuring the input resistance and the radiation resistance of a vertical and a horizontal dipole antenna, and then comparing the data with that obtained in the absence of the LHM will provide a quick assessment of the quality of the sample of the metamaterial. An attempt to "measure" the image of the antenna directly would require collecting and processing a three dimensional array of data and this is not an easy task.

The stored electromagnetic energy density in a metamaterial with a permittivity and permeability functions given by (1) and (2) is [23]

$$w = w_E + w_M, \tag{4}$$

where

$$w_E(t) = \frac{\varepsilon_0}{2} \boldsymbol{E}^2 + \frac{1}{\omega_p^2 \varepsilon_0} \left(\frac{\partial \boldsymbol{P}}{\partial t}\right)^2 \tag{5}$$

and

$$w_M(t) = \frac{\mu_0(1-F)\boldsymbol{H}^2}{2} + \frac{1}{2\omega_0^2 \mu_0 F} \left\{ \left(\frac{\partial \boldsymbol{M}}{\partial t} + \mu_0 F \frac{\partial \boldsymbol{H}}{\partial t}\right)^2 + \omega_0^2 (\boldsymbol{M} + \mu_0 F \boldsymbol{H})^2 \right\}. \tag{6}$$

are the electric and the magnetic parts of the energy density. In (5) and (6) $\boldsymbol{E}$, $\boldsymbol{H}$, $\boldsymbol{P}$ and $\boldsymbol{M}$ are the electric field, magnetic field, polarization and magnetization vectors, respectively.

In order to compare the amount of electromagnetic energy stored into the LHM disk, expressions (5) and (6) have been integrated over the volume of the disk for a radiating system Fig. 1(f) and a non-radiating system Fig. 1(e). The results are presented in Fig. 4. The energy stored in the LHM in the absence of a PMC, or PEC plate is also shown. The input power is the same in all three cases. Figure 4 shows that the amount of energy stored in the LHM is largest for a non-radiating system. This outcome is intuitively acceptable. The characteristic time of the formation of the "non-radiating state" is large in comparison with the period $T$ of the excitation frequency. The relatively large amount of stored energy is an indication that strong electromagnetic field can be created inside the LHM disk by using a non-radiating system. This suggests the possible use of a non-radiating system as a non-linear device. To illustrate this, an LHM disk with second-order nonlinearity is considered. It has been shown [15] that by connecting diodes to the split-ring resonators of the metamaterial an effective, second-order, non-linear magnetization is induced. Detailed analysis of the second harmonic generation process in a left-handed metamaterial has been presented [24]. Here a simple relation between the second-order nonlinear magnetization $M_\varphi^{(NL)}(t)$ and the magnetic field

$$M_\varphi^{(NL)}(t) = \mu_0 \chi^{(2)} H_\varphi^2(t) \qquad (7)$$

is used. Figure 5 compares the second harmonic generation efficiency for a non-radiating system "LHM and PMC", radiating system "LHM and PEC" and for a LHM disk irradiated by the same antenna without PEC, or PMC plate. The second-harmonic power depends upon the square of the input power, but the efficiencies of the conversion process are quite different for the three devices. The non-radiating system offers a clear advantage, giving the highest efficiency, as Fig. 5 shows. This can be associated with the strong non-propagating fields existing in the material in full accordance with Fig. 4. This suggests a possible application of a non-radiating configuration as a parametric amplifier.

The advantage in the conversion efficiency, provided by the non-radiating system, is illustrated in Fig. 6 where the magnetic field distributions, created by each of the systems considered are plotted. Note, that the non-radiating system generates the highest amount of second harmonic power although the amount of input – fundamental frequency power – is the lowest.

## 4. Conclusions

The perfect lens property of the LHMs has been exploited to couple a wire dipole antenna to its mirror image. Both PEC and PMC-mirrors have been considered. It is shown that with vertical antennae the PMC mirror results in a non-radiating configuration, whereas the PEC mirror stimulates radiation. The non-radiating system allows large amounts of electromagnetic energy to be stored in the metamaterial and this indicates that strong non-propagating electromagnetic field exists in this case. The latter suggests the use of non-radiating systems as non-linear devices. The advantage provided by such systems in a second-harmonic generation process is demonstrated.

**Acknowledgements**

This work is supported by the Engineering and Physical Sciences Research Council (UK) under the Adventure Fund Programme.

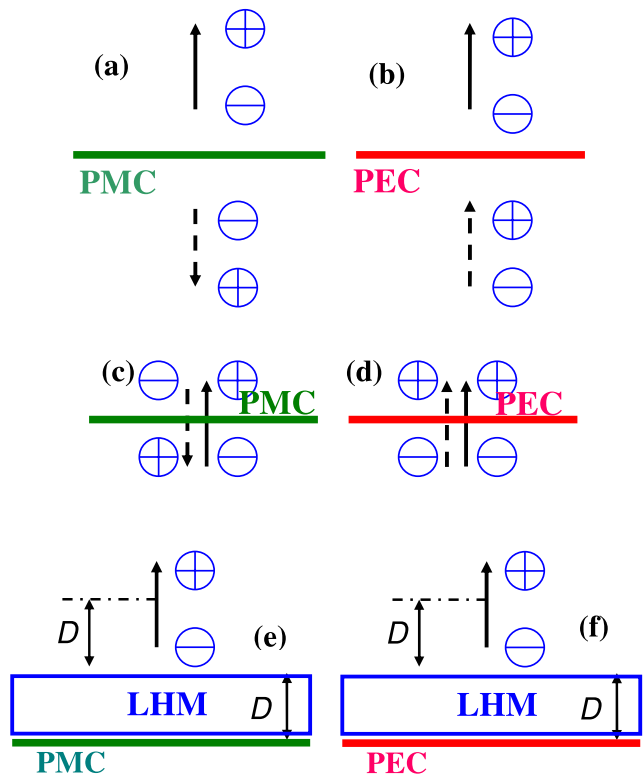

Figure 1

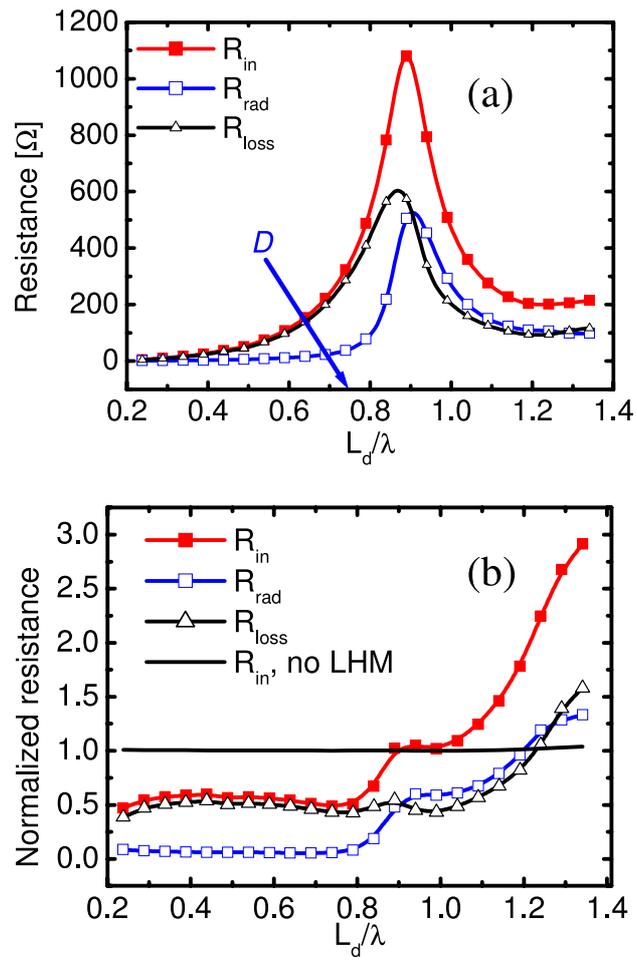

Figure 2

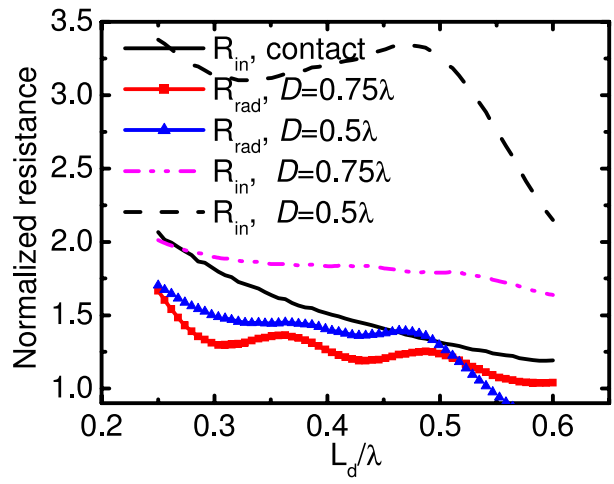

Figure 3

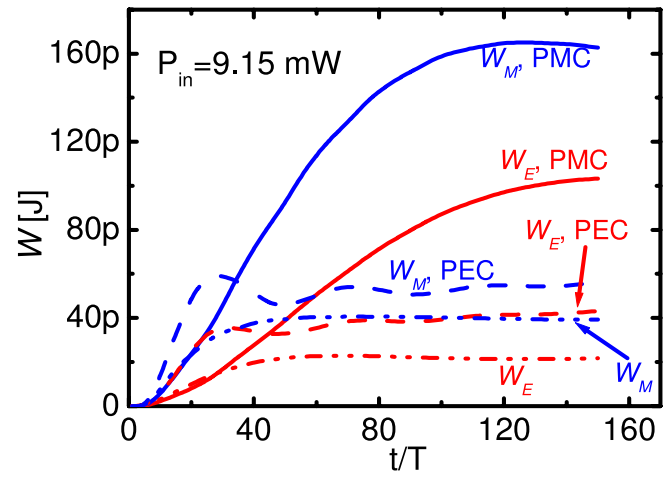

Figure 4

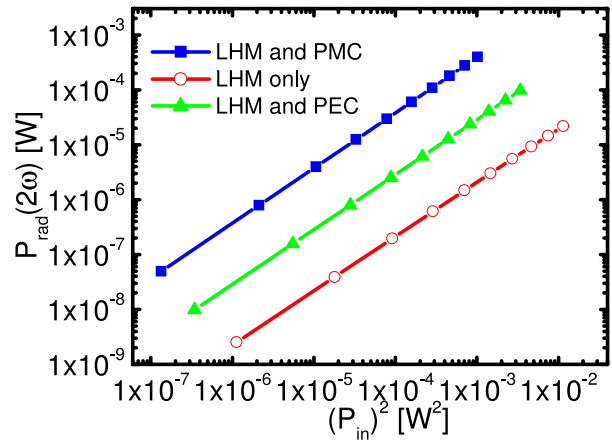

Figure 5

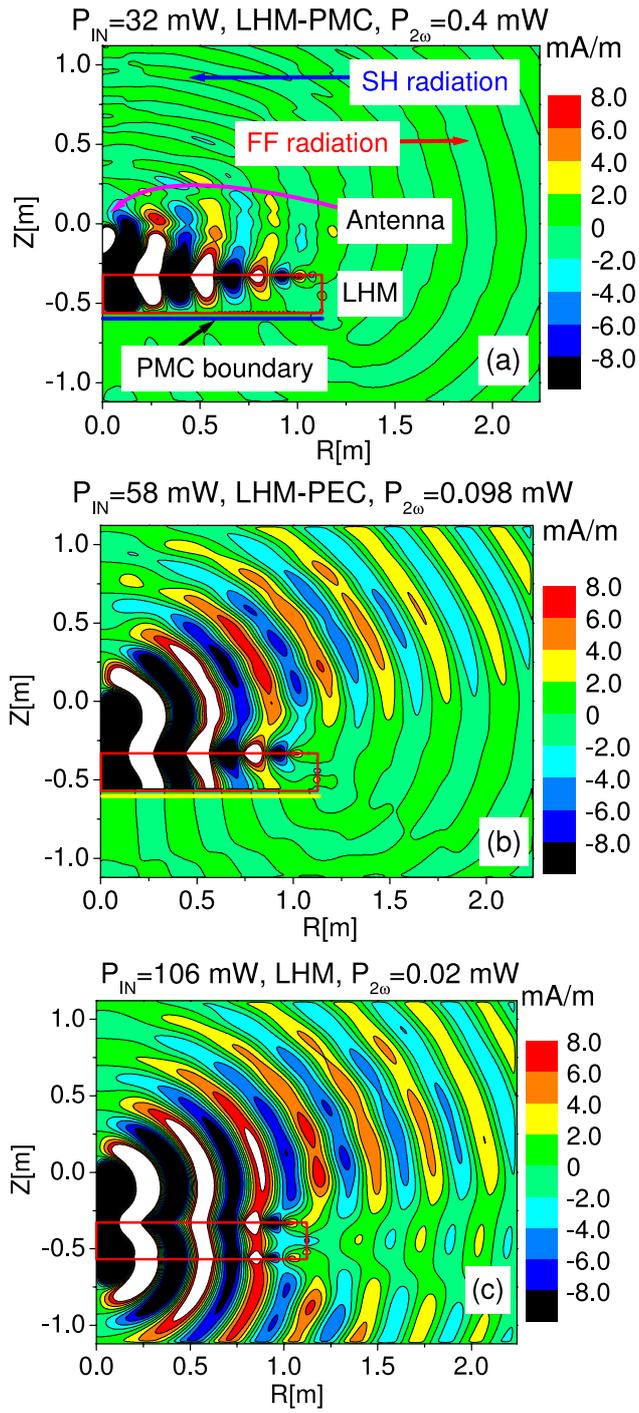

Figure 6

**Figure captions**

**Figure 1.** Non-radiating and radiating systems formed by coupling a vertical electric dipole to its mirror image. The images formed by a PMC- (a) or a PEC-mirror (b) are dipoles of the same magnitude: inverted (PMC) or non-inverted (PEC). Infinitesimal dipole on a PMC surface (c) does not radiate whereas the presence of a PEC surface (d) will double its radiation resistance. A slab made of an LHM of thickness $D$ with an index of refraction equal to -1 is used to superimpose a dipole of finite length onto its mirror image: (e) PMC and (f) PEC.

**Figure 2.** PMC case. Radiation $R_{rad}$, input $R_{in}$ and loss resistance $R_{loss}$ of the non-radiating configuration shown in Fig. 1(e) as a function of the antenna length. The excitation wavelength is kept fixed to $\lambda = 0.299\,m$. (a) Absolute values; (b) Normalized values. The curves are normalized to the input resistance of an antenna of the same length emitting in free space. The thickness of the LHM disk is $D = 0.75\lambda$. The curve [$R_{in}$, no LHM] shows the input resistance of an antenna in the absence of the LHM disk. The distance between the center of the antenna and the PMC-boundary is 2$D$. Note, that in this case the PMC-boundary has no impact on the antenna radiation resistance.

**Figure 3**. PEC case, normalized resistance. The curves are normalized to the input resistance of an antenna of the same length emitting in free space. Two different values of the disk thickness – $D = 0.75\lambda$ and $D = 0.5\lambda$ – are used. The curve labeled [$R_{in}$, contact] refers to a dipole with the end sitting directly on the PEC boundary in the absence of the LHM disk. The distance between the center of the dipole and the PEC is $D = L_D/2$ in this case.

**Figure 4.** Magnetic ($W_M$) and electric ($W_E$) parts of the electromagnetic energy stored in the LHM disk situated on a plate made of a PMC, ($W_M$,PMC;$W_E$,PMC) or a PEC ($W_M$,PEC;$W_E$,PEC). The corresponding result obtained in the absence of any plate ($W_M$; $W_E$) is also shown. The input power supplied to the antenna is one and the same in the three cases. $t$ is the time coordinate, $T$ is the period of excitation. $W_E$ and $W_M$ have been obtained by integrating $w_E$ and $w_M$, given by (5) and (6), respectively, over the volume of the disk.

**Figure 5**. Dependence of the second-harmonic radiated power on the input power supplied to the antenna for a LHM disk with second-order nonlinearity for three configurations; LHM disk on a PMC plate (Fig. 1(e)); LHM disk on a PEC plate (Fig. 1(f)); LHM disk alone. $L_D = 0.45\lambda$.

**Figure 6**. Azimuthal magnetic field component distributions $H_\varphi(R,Z)$ [mA/m] for the three configurations considered; (a) LHM disk on a PMC plate; (b) LHM disk on a PEC plate; (c) LHM disk alone. Note that the second-harmonic wave is easy to see in the case of a non-radiating system (Fig 6(a)) despite the fact that the input power is the lowest in this case. FF fundamental frequency, SH second harmonic.